\documentclass[11pt,a4paper]{article} 
\usepackage{caption}         %
\usepackage{charter}         
\usepackage{graphicx}        
\usepackage{amssymb,amsmath} 
\usepackage{multicol}        
\usepackage{url}             
\usepackage{enumitem}        
\usepackage{marvosym}        
\usepackage{wrapfig}         
\usepackage[T1]{fontenc}     
\usepackage{mathptmx}        
\usepackage{datetime}        
\usepackage{fancyhdr}        
\newdateformat{mydate}{\monthname[\THEMONTH] \THEYEAR}          
\usepackage[pdfpagemode=FullScreen, colorlinks=false]{hyperref} 
\usepackage{wrapfig}
\usepackage{subcaption}
\usepackage{xcolor}

\newcommand{\HorRule}[1]{\noindent\rule{\linewidth}{#1}}

\newcommand{\LetterTitle}[1]
{
    \begin{center}
    \usefont{T1}{ptm}{b}{n}  
    \Huge #1
    \end{center}	
    \par \normalsize \normalfont
    \HorRule{3pt}
}

\newcommand{\LetterHeader}[3]
{
    \usefont{T1}{ptm}{n}{n}  
    \hfill \textsc{Volume #1, Issue #2, #3}  
    \par \normalsize \normalfont
}

\newcommand{\SectionTitle}[1]
{
    \HorRule{1pt}
    \usefont{T1}{ptm}{b}{n}               
    \vspace{3pt}\Large #1\vspace{1pt}    
    \par \normalsize \normalfont
    \HorRule{1pt}
}

\newcommand{\NewsTitle}[1]
{
    \begin{center}
    \usefont{T1}{ptm}{b}{n}               
    \vspace{14pt}\Large #1\vspace{1pt}    
    \par \normalsize \normalfont
    \end{center}
}

\newcommand{\NewsAuthor}[1]
{
    \begin{center}
    \usefont{T1}{ptm}{n}{n}               
    \textsc{#1} \vspace{4pt} 
    \par \normalfont
    \end{center}
}

\graphicspath{{./figs/}}

\begin{document}

\LetterHeader{1}{1}{February 1, 2016}  
\LetterTitle{TC-CCPS Newsletter}       
\clearpage

\SectionTitle{Technical Articles}

\NewsTitle{Multi-Modal Attack Detection for Cyber-Physical Additive Manufacturing}
\NewsAuthor{Shih-Yuan Yu, Arnav Vaibhav Malawade, Mohammad Abdullah Al Faruque}
\NewsAuthor{Department of Electrical Engineering and Computer Science, University of California, Irvine}

\section{Introduction}
\label{sec:introduction}
Cyber-Physical Additive Manufacturing (AM) constructs a physical 3D object layer-by-layer according to its digital representation and has been vastly applied to fast prototyping and the manufacturing of functional end-products across fields.
The computerization of traditional production processes propels these technological advancements; however, this also introduces new vulnerabilities, necessitating the study of cyberattacks on these systems~\cite{yampolskiy2018security}. 
The AM Sabotage Attack is one kind of kinetic cyberattack that originates from the cyber domain and can eventually lead to physical damage, injury, or even death~\cite{applegate2013dawn}. 
By introducing inconspicuous yet damaging alterations in any specific process of the AM digital process chain, the attackers can compromise the structural integrity of a manufactured component in a manner that is invisible to a human observer. 
If the manufactured objects are critical for their system, those attacks can even compromise the whole system's structural integrity and pose a severe safety risk to its users. 
For example, an inconspicuous void (less than 1 mm in dimension) placed in the 3D design of a tensile test specimen can reduce its yield load by 14\% \cite{sturm2014cyber}. 
However, security studies primarily focus on securing digital assets~\cite{holland2018intellectual}, overlooking the fact that AM systems are CPSs.

The AM system, or the printer, is comprised of a set of connected hardware components, and thus can unintentionally produce analog emissions during the operation of printing through different physical side-channels such as acoustics, electromagnetic radiation, vibration, and power.
In AM systems, the information flow in the cyber domain has at least one corresponding control signal sent to the physical domain.
This signal flow, in turn, actuates the physical processes accordingly, resulting in side-channel emissions that have a high degree of mutual information with the digital control signals. 
This property allows our group to a series of research on utilizing the correlation between the two domains and validating that physical domain signals match their cyber domain counterparts~\cite{chhetri2016kcad, AbdullahAlFaruque2016, al2016forensics, chhetri2017fix, chhetri2017security, chhetri2017side, faezi2019oligo, chhetri2018information}. 
Sabotaging the structural integrity of 3D objects requires the attacker to make subtle variations to one or more of the sub-processes in the AM process chain, resulting in a change in the printer's control parameters and a corresponding change to its analog emissions.
In this case, monitoring the operation of the targeted AM systems can be the most direct defense~\cite{chhetri2016kcad}. 
To detect such modifications, our group proposes an attack detection system that continuously monitors and analyzes the different side-channel information leaked during the operation of AM systems, allowing us to identify unusual analog emissions resulting from potential sabotage attacks.
\section{The AM Process Chain}
\label{sec:background}
\begin{figure}[h!]
    \includegraphics[width=1.0\linewidth]{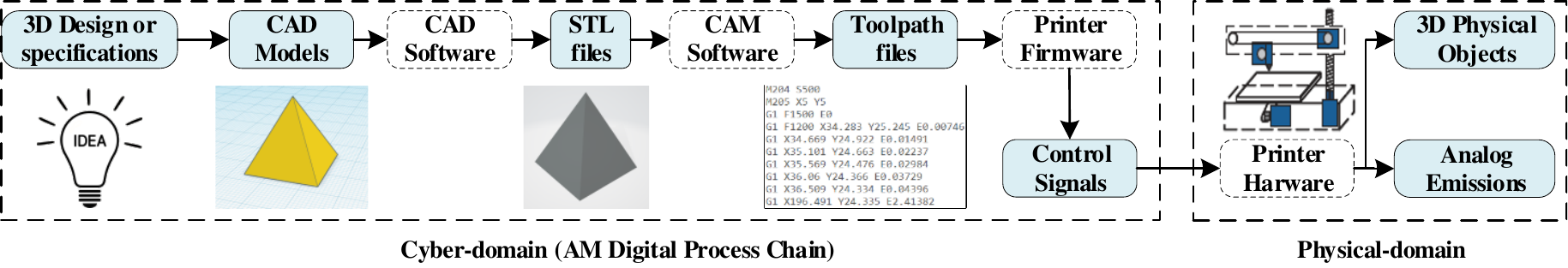}
    \caption{The AM process chain.}
    \label{fig:process_chain}
\end{figure}

The manufacturing of 3D objects in AM requires a chain of cyber-physical processes~\cite{yampolskiy2018security}.
We primarily describe the process chain for the Fused Deposition Model (FDM) based AM. 
As shown in Figure~\ref{fig:process_chain}, the process chain begins in the cyber domain with an idea for a 3D object.
The first item on the workflow is to substantiate the design specifications using Computer-Aided Design (CAD) tools.
Next, the generated CAD model gets converted into the STereoLithography (STL) format that uses a series of triangles to model the surface geometry.
In the Computer-Aided Manufacturing (CAM) process, the slicing algorithms convert the STL files into a layer-by-layer description file which instructs the printer on how to create the 3D object. 
The description file uses a Numerical Control Programming Language called G/M-code.
The G-codes describe the motion and flow settings of the printer, determining the speed of the nozzle along each axis and the amount of material to deposit at each printing step. 
As an example, \textit{G1 F2100 X5 Y6 Z1.2 E2.1} represents a single line of G-code for controlling the movement of the nozzle, where G1 means coordinated linear motion, F defines travel feed rate (speed) which is measured in mm/min, and distance and extrusion are measured in mm. 
On the other hand, the M-codes control the machine settings such as temperature, coolant.
Lastly, the printer's firmware interprets each received G/M-code instruction into the corresponding control signals to actuate the printer hardware and, in turn, print the object.

In the physical domain, the FDM-based printer uses thermoplastic filament materials to form 3D physical objects.
The printer uses motors and belts to control the heated nozzle to melt the materials, depositing the extrusion onto the surface of the build plate at a precisely controlled rate.
An array of stepper motors enables a fine-grained degree of control, each controlling movement along a single axis.
These components act as a significant source of information leakage because of their analog emissions during printer operation.
The analog emissions come from various side-channels of the printer (vibration, acoustic, magnetic, and power side-channels), which are streams of information outside of the primary data path containing information which can be correlated with the data in the primary path.
In this case, the primary data path includes the control signals sent to the printer. 
The stepper motors are a major source of vibration and acoustic emissions due to the fluctuating radial force and torque ripples working on the stator core of the stepper motor. 
The varying electric field in the stepper motors also leaks data to the magnetic side channel.
The power side-channel reveals the primary consumers of power in a 3D printer, such as the heating elements, stepper motors, fans, bed heater, and the internal control circuitry.
In short, through these side-channels, the unintentional information leakage about the primary data path enables the inference of control signal values from side-channel information.
\section{Sabotage Attack Detection for AM Systems}
As for the threat model, the attackers can exploit multiple attack surfaces on the AM process chain (Figure~\ref{fig:process_chain}) to sabotage manufactured parts. 
From the cyber domain, the adversaries can modify the intermediate forms of the 3D object by altering the integrity of the tools, firmware, or computers in the process chain, eventually resulting in the modification of the printer's control signals and the sabotage of the printed object. 
Cyber-security techniques can help mitigate risks further upstream in the digital process chain; however, AM systems often have network connectivity or physical access ports that can enable an adversary to exploit firmware vulnerabilities and compromise the system. 
To this end, we focus our on detecting adversarial attacks to the printer firmware that modify the control parameters directly. 
With this in mind, we assume that the G-codes sent to the AM system have not been previously tampered with and are representative of the correct 3D object model. 
In this case, the movement of the AM machine's internal components during a sabotage attack will not match the movements described by the instructions given in the G-code file, and thus, will result in different side-channel emissions.
By utilizing these side-channel emissions with the G-codes sent to the printer, we can infer the values of the control signals and then determine if the printer firmware has been hacked.

\begin{wrapfigure}{r}{.45\textwidth}
    \vspace{1.0em}
    \includegraphics[width=\linewidth]{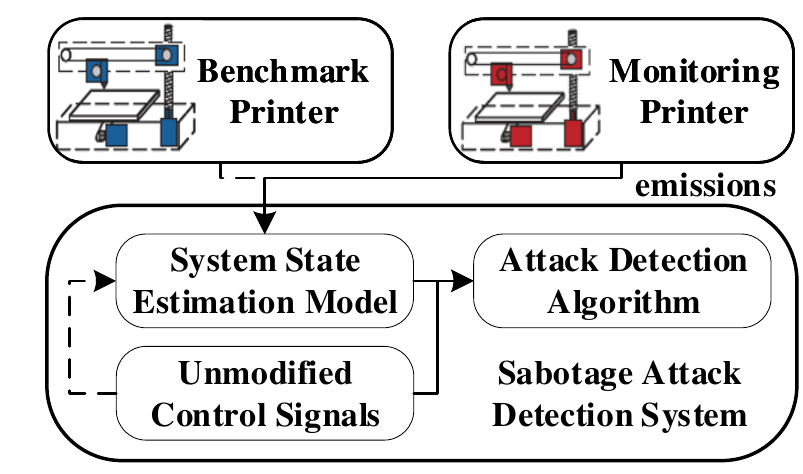}
    \caption{System Architecture.}
    \label{fig:archi}
\end{wrapfigure}

The architecture of our proposed Sabotage Attack Detection System is described in Figure~\ref{fig:archi}.
To infer the control signals, the \textbf{System State Estimation Model} learns the relationship between the various side-channel emissions and the various control signals using supervised machine learning approaches.
The dataset for training consists of the analog emissions from the \textbf{Benchmark Printer} along with their corresponding control signals, which are parsed from the G-codes. 
Once the mapping between analog emissions and control signals has been learned, our system first infers each control signal by continuously observing the analog emissions from the \textbf{Monitoring Printer}.
The \textbf{Attack Detection Algorithm} then compares the unmodified control signals to the inferred signals to determine if a sabotage attack has occurred.
\section{Results}
\label{sec:results}
\begin{wrapfigure}{r}{0.25\textwidth}
    \vspace{-1.5em}
    \includegraphics[width=\linewidth]{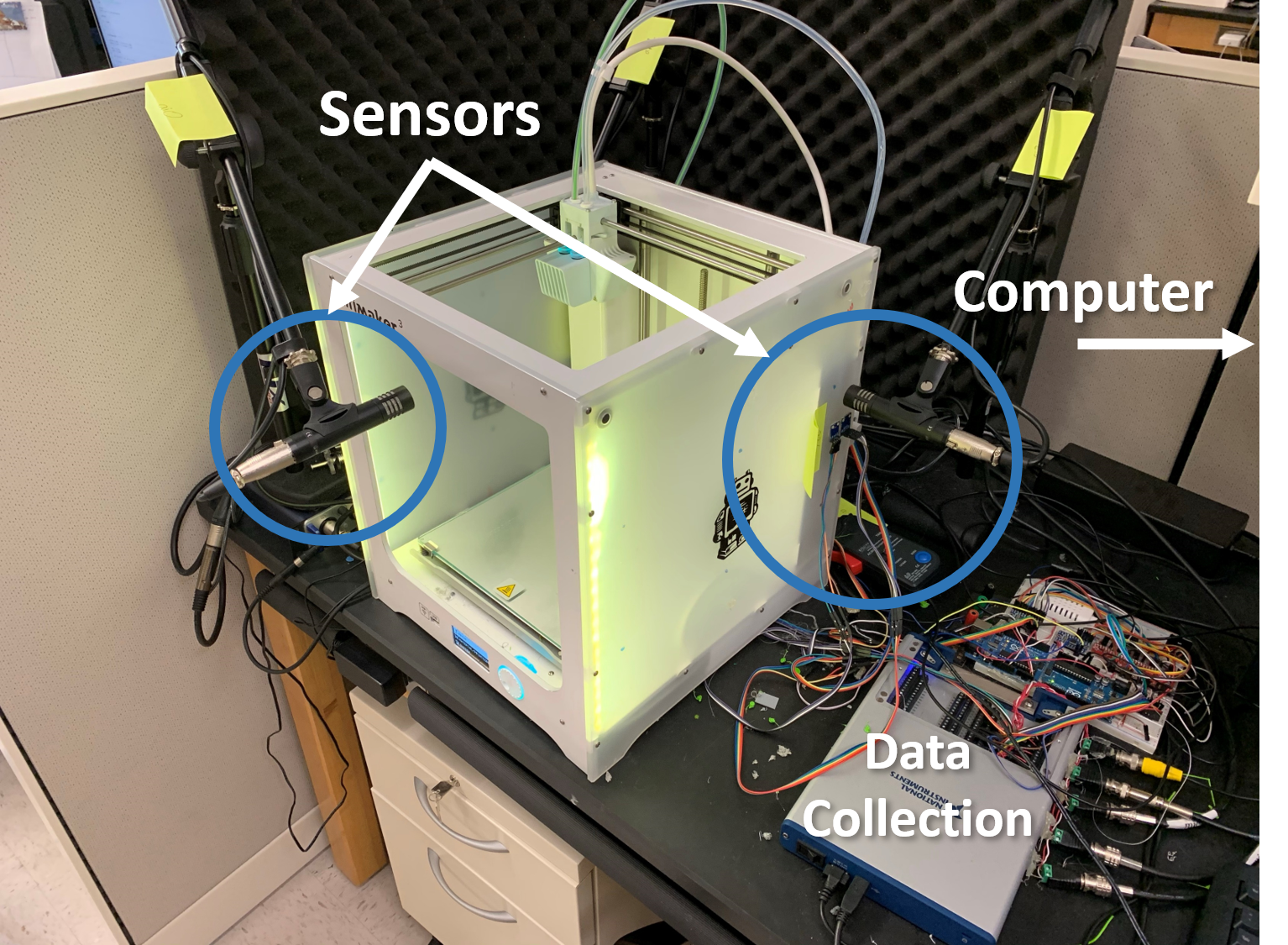}
    \caption{Experimental setup}
    \label{fig:testbed}
    \vspace{-3em}
\end{wrapfigure}
Our experimental setup is shown in Figure~\ref{fig:testbed}, consisting of an Ultimaker 3 3D-printer, four microphones, three vibration sensors (accelerometers), three magnetic sensors (magnetometers), and one current sensor. 
We use timestamps to match each G-code to its corresponding side-channel emissions recorded from the sensors.
To evaluate the performance of our proposed detection system, we combined data from several different 3D-prints to produce a 20-gigabyte dataset containing 60,959 rows with 18,276 features per row. 
In the following sections, the selected results for state estimation and attack detection are shown (see~\cite{8984311} for more results). 

\subsection{AM System State Estimation}
\label{subsec:stateestimation}
\begin{figure}[htb]
\begin{subfigure}{0.5\textwidth}
    \centering
    \includegraphics[width=\linewidth]{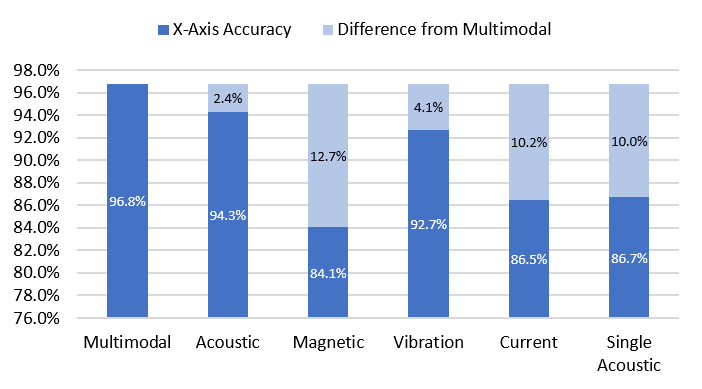}
\end{subfigure}
\begin{subfigure}{0.5\textwidth}
    \centering
    \includegraphics[width=\linewidth]{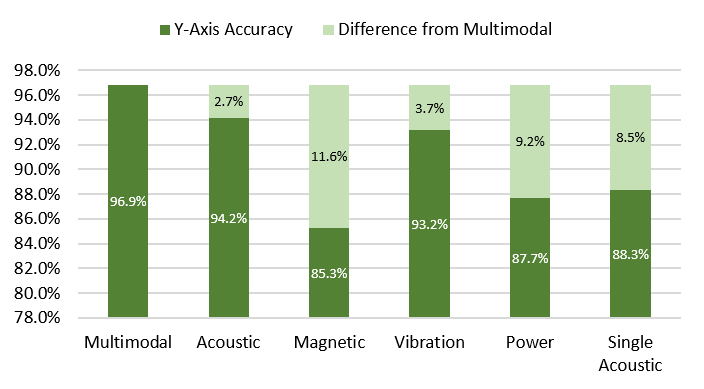}
\end{subfigure}
    \caption{The comparison of multimodal approach and unimodal approaches in axis movement prediction.}
    \label{fig:modality-accuracy-comparison}
\end{figure}

The ability of our system to detect sabotage attacks depends on how precisely our system models the relationship between side-channel information and the corresponding machine control parameters.
To improve our system's performance, we use multiple modalities of side-channel data for control signal estimation. To evaluate our multi-modal approach in comparison to single side-channel (unimodal) methods, we assessed the movement-axis prediction ($A_x$ and $A_y$) performance of our classifiers using data from each modality. The results of the highest accuracy classifiers in each modality in comparison to our multi-modal approach and the single acoustic sensor approach presented in \cite{chhetri2016kcad} are shown in Figure \ref{fig:modality-accuracy-comparison}. 
As shown in the figure, our multi-modal technique outperforms the uni-modal methods as well as the technique from \cite{chhetri2016kcad}. Notably, the acoustic and vibration modalities show the highest uni-modal accuracy and reach within 5\% of the accuracy of the multi-modal technique. Overall, these results demonstrate that our multi-modal technique results in performance improvement over uni-modal methods.

\subsection{AM Sabotage Attack Detection}
\label{subsec:realworld}
To evaluate our attack detection performance, we generate synthetic attacks by adding adversarial modifications to G-Codes.
Then, we pass the unmodified sensor data and the modified G-code files as inputs to our attack detection system to evaluate its ability to detect mismatches.
Our overall row-level accuracy is \textbf{99.17\%} for movement-axis prediction.
For axis-velocity prediction our average accuracy is \textbf{96.64\%}. 
Our overall attack detection accuracy for all control parameters is \textbf{98.15\%}.
Although these attacks are synthetic, they provide empirical benchmarks to demonstrate the attack detection system's capabilities. 
\begin{figure}[ht!]
\centering
\begin{subfigure}{0.5\textwidth}
\centering
    \begin{subfigure}{0.7\textwidth}
        \includegraphics[width=\linewidth]{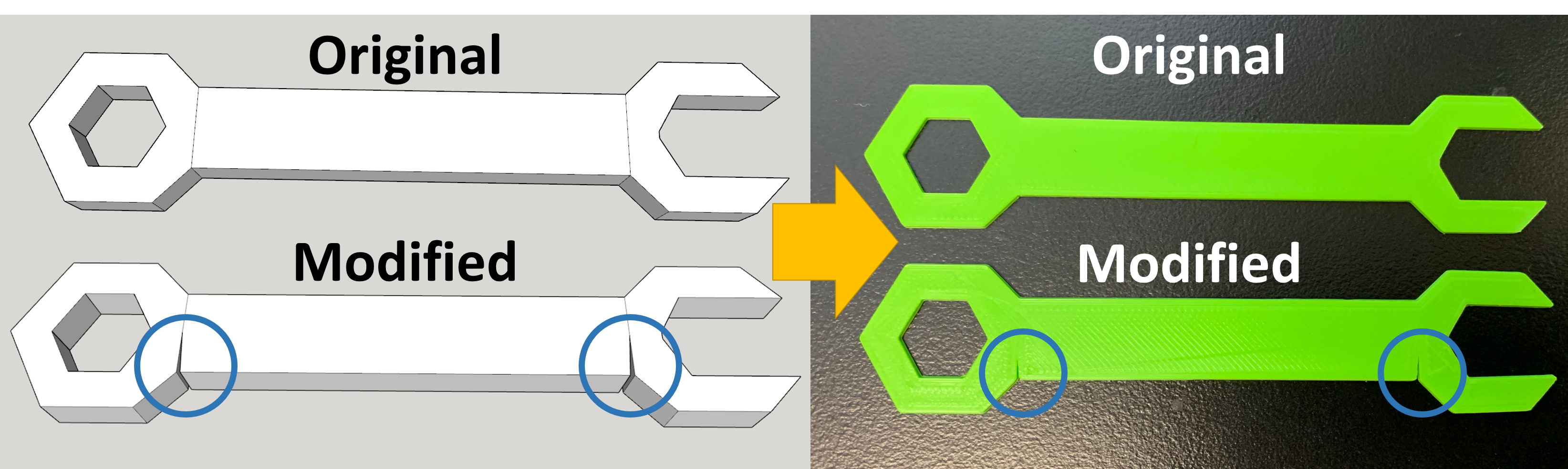}
        \caption{Original and modified wrench.}
        \label{fig:adversarially_modified_wrench}
    \end{subfigure}%
    
    \begin{subfigure}{0.7\textwidth}
        \includegraphics[width=\linewidth]{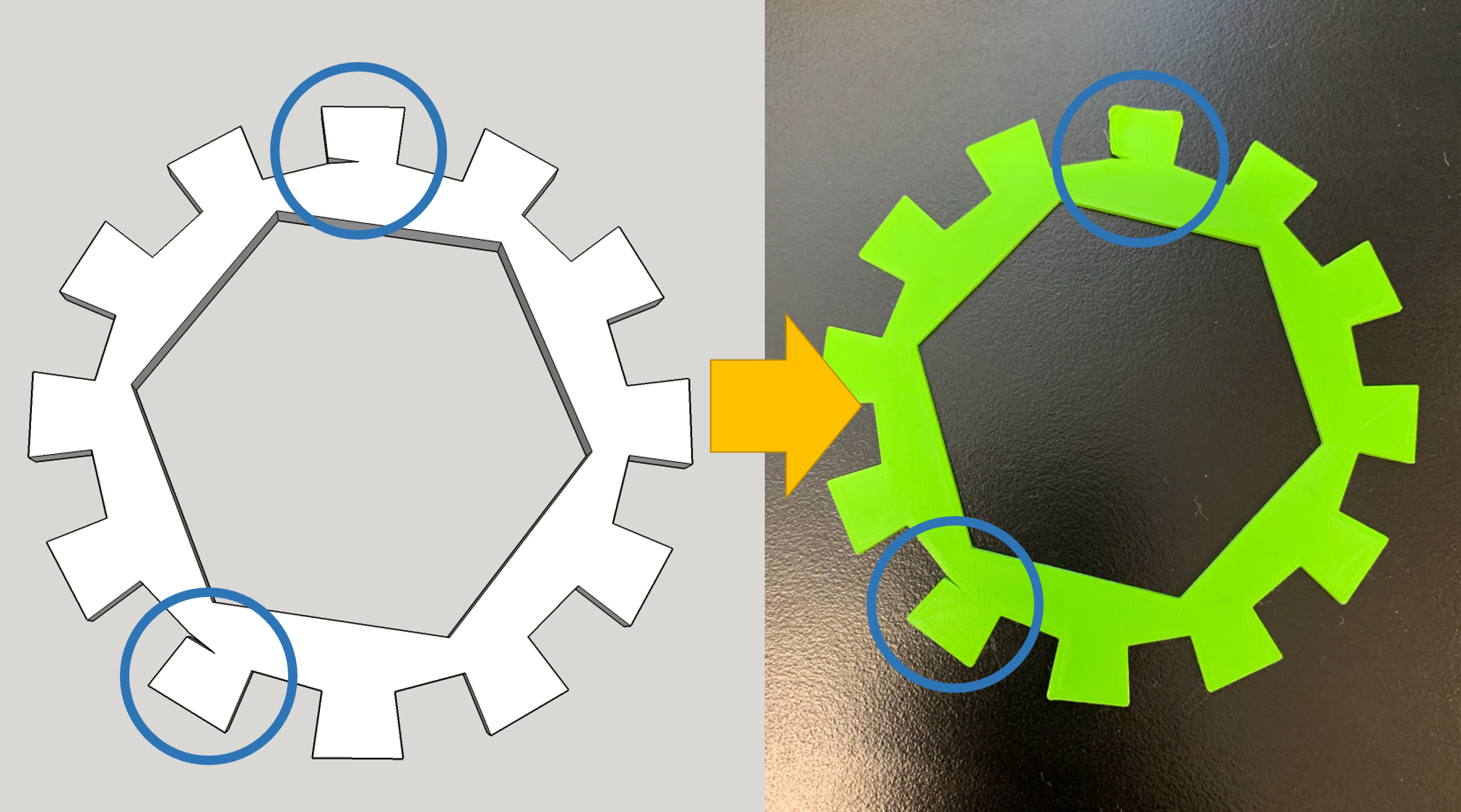}
        \caption{Original and modified gear.}
        \label{fig:adversarially_attacked_gear}
    \end{subfigure}
\end{subfigure}%
\begin{subfigure}{0.5\textwidth}
\centering
    \includegraphics[width=0.7\linewidth]{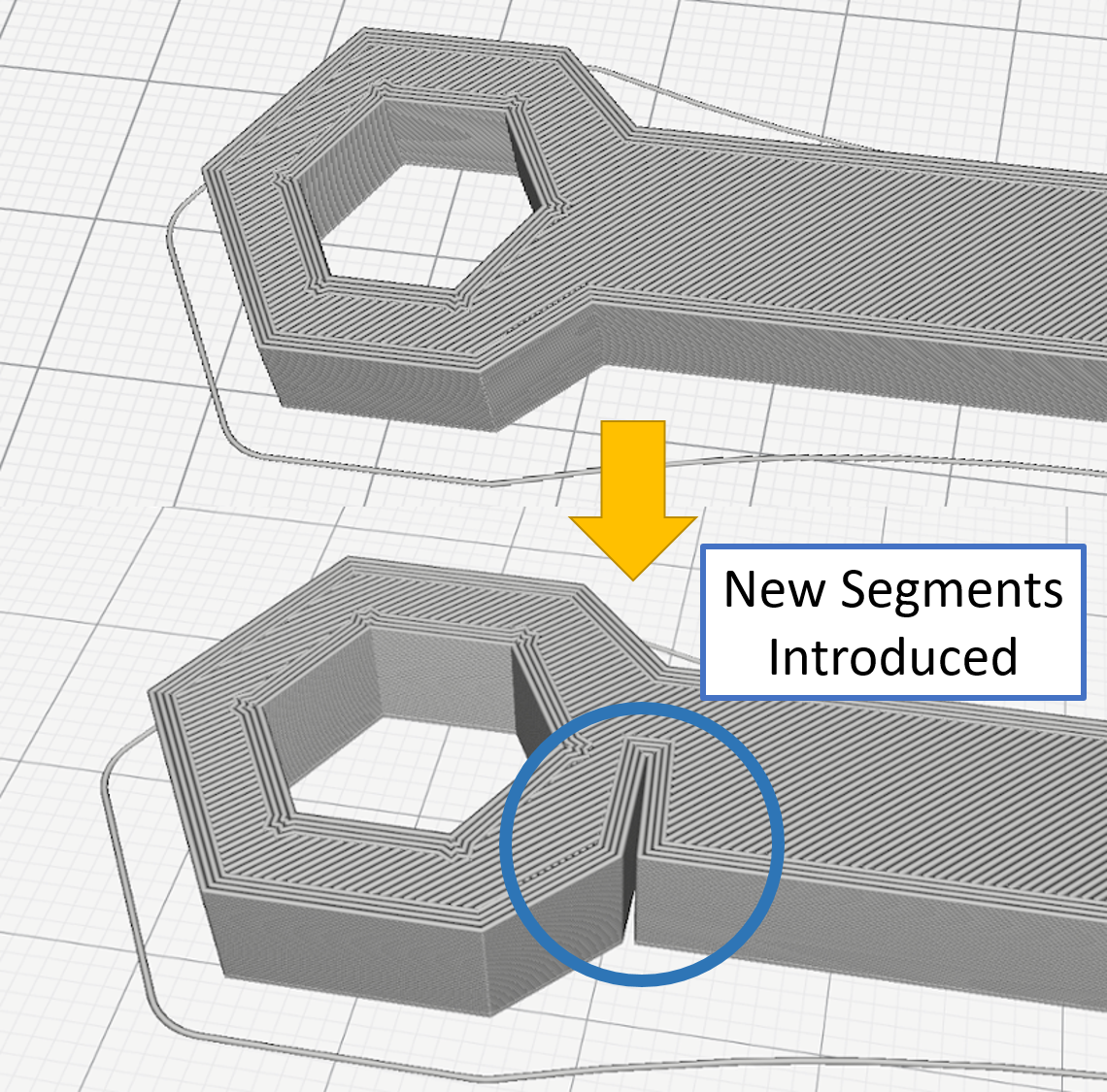}
    \caption{Adversarial modifications in details.}
    \label{fig:adv_segments}
\end{subfigure}

\caption{The adversarially modified objects used for real-world testing.}
\end{figure}

To validate these results in a real scenario, we tested two real-world sabotage attacks: a gear and a wrench. 
Since gears and wrenches are generally placed in high-stress scenarios, it is plausible that a kinetic cyberattack could compromise their structural integrity and significantly increase the chance of equipment failure. 
Moreover, these parts serve as analogues to other load-bearing 3D printed parts to demonstrate that our detection system can identify kinetic cyberattacks on these types of components.
The original and attacked wrenches are shown in Figure~\ref{fig:adversarially_modified_wrench}, and the adversarial modifications to the gear are shown in Figure~\ref{fig:adversarially_attacked_gear}. 
Although these specific modifications are visible to the human eye in the final print, this was done primarily for demonstrative purposes. 
These modifications can easily be concealed in internal layers of the print without modifying the outward appearance of the object, enabling it to pass visual quality inspections done by both machines and employees. 
For both test cases, the detection system was able to detect the attack successfully. These tests demonstrate that our methodology can be applied to real-world, practical AM tasks. 
\section{Conclusion}\label{sec:conclusion}
AM systems are core to the future of manufacturing and Industry 4.0. 
Since these systems are advancing in terms of complexity and connectivity as well as producing parts for increasingly critical and taxing applications, the security of these systems is a significant concern. 
Since AM systems are CPSs, they present a broad and diverse attack surface, making attack detection and prevention a daunting task. 
Our proposed sabotage attack detection system has demonstrated the ability to detect various attacks by correlating multiple forms of physical-domain emissions of the AM system with cyber-domain information.
We have achieved an overall detection accuracy of 98.15\% on synthetic benchmarks and demonstrated that our system could detect two real-world scenarios of sabotage attacks. 
Although our solution addresses a significant security risk in AM systems, more work is needed to address the broad scope of potential AM vulnerabilities.

\bibliographystyle{abbrv}
\bibliography{kcad} 

\begin{thebibliography}{10}

\bibitem{al2016forensics}
M.~A. Al~Faruque, S.~R. Chhetri, A.~Canedo, and J.~Wan.
\newblock Forensics of thermal side-channel in additive manufacturing systems.
\newblock {\em University of California, Irvine}, 2016.

\bibitem{applegate2013dawn}
S.~D. Applegate.
\newblock The dawn of kinetic cyber.
\newblock In {\em 2013 5th international conference on cyber conflict (CYCON
  2013)}, pages 1--15. IEEE, 2013.

\bibitem{chhetri2017side}
S.~R. Chhetri and M.~A. Al~Faruque.
\newblock Side channels of cyber-physical systems: Case study in additive
  manufacturing.
\newblock {\em IEEE Design \& Test}, 34(4):18--25, 2017.

\bibitem{chhetri2016kcad}
S.~R. Chhetri, A.~Canedo, and M.~A.~A. Faruque.
\newblock Kcad: kinetic cyber-attack detection method for cyber-physical
  additive manufacturing systems.
\newblock In {\em Proceedings of the 35th international conference on
  Computer-Aided Design}, page~74. ACM, 2016.

\bibitem{chhetri2018information}
S.~R. Chhetri, S.~Faezi, and M.~A. Al~Faruque.
\newblock Information leakage-aware computer-aided cyber-physical
  manufacturing.
\newblock {\em IEEE Transactions on Information Forensics and Security},
  13(9):2333--2344, 2018.

\bibitem{chhetri2017fix}
S.~R. Chhetri, S.~Faezi, and M.~A.~A. Faruque.
\newblock Fix the leak!: an information leakage aware secured cyber-physical
  manufacturing system.
\newblock In {\em Proceedings of the Conference on Design, Automation \& Test
  in Europe}, pages 1412--1417. European Design and Automation Association,
  2017.

\bibitem{chhetri2017security}
S.~R. Chhetri, N.~Rashid, S.~Faezi, and M.~A. Al~Faruque.
\newblock Security trends and advances in manufacturing systems in the era of
  industry 4.0.
\newblock In {\em 2017 IEEE/ACM International Conference on Computer-Aided
  Design (ICCAD)}, pages 1039--1046. IEEE, 2017.

\bibitem{faezi2019oligo}
S.~Faezi, S.~R. Chhetri, A.~V. Malawade, J.~C. Chaput, W.~H. Grover, P.~Brisk,
  and M.~A. Al~Faruque.
\newblock Oligo-snoop: A non-invasive side channel attack against dna synthesis
  machines.
\newblock In {\em NDSS}, 2019.

\bibitem{AbdullahAlFaruque2016}
A.~Faruque, M.~Abdullah, S.~R. Chhetri, A.~Canedo, and J.~Wan.
\newblock Acoustic side-channel attacks on additive manufacturing systems.
\newblock In {\em Proceedings of the 7th International Conference on
  Cyber-Physical Systems}, page~19. IEEE Press, 2016.

\bibitem{holland2018intellectual}
M.~Holland, J.~Stjepandi{\'c}, and C.~Nigischer.
\newblock Intellectual property protection of 3d print supply chain with
  blockchain technology.
\newblock In {\em 2018 IEEE International Conference on Engineering, Technology
  and Innovation (ICE/ITMC)}, pages 1--8. IEEE, 2018.

\bibitem{sturm2014cyber}
L.~Sturm, C.~Williams, J.~Camelio, J.~White, and R.~Parker.
\newblock Cyber-physical vunerabilities in additive manufacturing systems.
\newblock {\em Context}, 7(2014):8, 2014.

\bibitem{yampolskiy2018security}
M.~Yampolskiy, W.~E. King, J.~Gatlin, S.~Belikovetsky, A.~Brown, A.~Skjellum,
  and Y.~Elovici.
\newblock Security of additive manufacturing: Attack taxonomy and survey.
\newblock {\em Additive Manufacturing}, 21:431--457, 2018.

\bibitem{8984311}
S.~{Yu}, A.~V. {Malawade}, S.~R. {Chhetri}, and M.~A. {Al Faruque}.
\newblock Sabotage attack detection for additive manufacturing systems.
\newblock {\em IEEE Access}, 8:27218--27231, 2020.

\end{thebibliography}

\end{document}